\documentclass[preprint]{aastex}
\usepackage{epsfig}

\newcommand{\beq}{\begin{equation}}
\newcommand{\eeq}{\end{equation}}
\newcommand{\beqa}{\begin{eqnarray}}
\newcommand{\eeqa}{\end{eqnarray}}
\newcommand{\msun}{\hbox {$M_{\sun}$ }}

\newcommand{\dvbto}{\hbox {$\Delta \rm V^{BTO}_{HB}$}}

\newcommand{\mvrr}{\hbox {$\rm M_v(RR)$}}

\newcommand{\afe}{\hbox{$ [\alpha/{\rm Fe}]$}}
\newcommand{\feh}{\hbox{$ [{\rm Fe}/{\rm H}]$}}
\newcommand{\ebv}{\hbox {${\rm E(\bv)}$}}
\newcommand{\dmv}{\hbox{$(m\!-\!M)_{V}$}}

\newcommand{\vi}{\hbox{$V\!-\!I$}}

\newcommand{\evi}{\hbox {${\rm E(\vi)}$}}

\begin{document}

\title{The Relative Age of the Thin and Thick Galactic Disks}
\author{Wilson M.\ Liu\altaffilmark{1} and Brian Chaboyer}
\affil{Department of Physics and Astronomy, Dartmouth College
6127 Wilder Laboratory, Hanover, NH, USA 03755-3528}
\altaffiltext{1}{Present Address: Steward Observatory, University of Arizona, Tucson AZ 85721}
\email{Brian.Chaboyer@Dartmouth.edu}

\begin{abstract}
We determine the relative ages of the open cluster NGC 188 and
selected Hipparcos field stars by isochrone fitting, and compare them
to the age of the thick disk globular cluster 47 Tuc.  The best fit
age for NGC 188 was determined to be 6.5 $\pm$1.0 Gyr.  The solar
metallicity Hipparcos field stars yielded a slightly older thin disk
age, 7.5 $\pm$0.7 Gyr.  Two slightly metal-poor ($\feh = -0.22$) field
stars whose kinematic and orbital parameters indicate that they are
members of the thin disk were found to have an age of $9.7\pm 0.6$
Gyr.  The age for 47 Tuc was determined to be 12.5 $\pm$1.5 Gyr.  All
errors are internal errors due to the uncertainty in the values of
metallicity and reddening.  Thus, the oldest stars dated in the thin
disk are found to be $2.8\pm 1.6$ Gyr younger than 47 Tuc.
Furthermore, as discussed by \citet{Chb99} 47 Tuc has a similar age
to three globular clusters located in the inner part of the Galactic
halo, implying that star formation in the thin disk started within
$2.8\pm 1.6$ Gyr of star formation in the halo.
\end{abstract}

\keywords{Galaxy: formation --- globular
clusters: general --- globular clusters: individual (47 Tuc)
open clusters and associations: individual (NGC 188)}

\section{Introduction}
The dating of the oldest stars in the Milky Way allows us to infer
the early history of star formation in our Galaxy, and hence, provides
us with important information regarding the formation of the Milky
Way.  A great deal of attention has been paid to the ages of globular
clusters (see for example the reviews by \cite{stetson,sara})
in the halo and there have been a number of studies which
determine the ages of the oldest open clusters in the thin disk 
(e.g.\  \citet{Car99}; \citet[ hereafter CGL]{Cha99}).
Typically, there appears to be a gap of several Gyr between the ages
of the globular clusters and the ages determined for the oldest stars
in the thin disk.  We have elected to investigate this further, in
order to quantify the age difference (if any) between the thin disk,
the thick disk and the halo of the Milky Way.  

In recent years, numerous techniques have been employed to determine
the ages of the oldest stars in the Galactic (thin)
disk. \citet{Jim98} examined the color magnitude diagram (CMD) of
Hipparcos field stars and arrived at a minimum disk age of 8 Gyr;
\citet{Car99} used isochrone fits to determine the ages of several old
open clusters.  \citet{Berg97} and \citet{Knox99} used observations of
local white dwarfs and theoretical white dwarf cooling curves to
determine a local disk age of $6.5-10$ Gyr and $9-13$ Gyr
respectively. \citet{Osw96} also used observations of white dwarfs to
place a minimum age of 9.5 Gyr on the galactic disk.  CGL studied the
old open cluster NGC 6791 and determined an age of $8.0\pm 0.5$ Gyr.

Globular clusters in the thick disk and halo have been the subject of
a number of recent investigations. For example, \citet{Sal98}
determined the ages of three thick disk globular clusters, 47 Tuc,
M71, and NGC 6352, and arrived at an age of 9.2 Gyr for all three.
\citet{Har97} examined two halo clusters, NGC 2419 and M92, in order
to determine their relative and absolute ages.  Their study determined
that the clusters had the same relative age to within 1 Gyr, and an
absolute age of $14-15$ Gyr.

In this paper, the relative age of the thin and thick disks is
measured, comparing the open clusters NGC 188, NGC 6791 and selected
Hipparcos field stars to the thick disk globular cluster 47 Tuc.  The
results of this analysis are related to the results of 
\citet[ hereafter CSA]{Chb99}, in which
the relative age of 47 Tuc to three halo clusters was determined using
the same input physics as this study.  Compared to other studies which
have investigated the relative age of the thin disk and the halo 
\citep[ eg.\ ]{Car99} this study has the advantage of using the same
assumptions and methodologies in dating the thin and thick disks and
the halo.  This allows us to determine precise relative ages and to
provide a reliable estimate of the error in the relative age
estimates.

Section \ref{secmodel} of this paper will discuss the stellar model
and isochrones used in this study.  Section \ref{thindisk} details the age
estimates for NGC 188 and the Hipparcos field stars (thin disk).  The
age of 47 Tuc (thick disk) is determined in \S \ref{sec47tuc}.  The results are
summarized in \S \ref{secsumm} where the relative age between the the thin and
thick disks and the halo is discussed.

\section{Stellar Models and Isochrones \label{secmodel}} 
Stellar evolution tracks were constructed using Chaboyer's  stellar
evolution code (CGL) for a range of masses of $0.50 - 1.30 \,\msun$ in
increments of 0.05 \msun.  The input physics, including low and high
temperature opacities, nuclear reaction rates, helium diffusion
coefficients and the equation of state were identical to those
described in CSA.  The models were evolved in 6000 time steps from the
zero age main sequence through the red giant branch.

The values of $\feh$ for each of set of stellar models were chosen to
reflect the observed values in each of the clusters.  To determine the
uncertainty in the age due to the uncertainty in \feh, stellar models
were calculated for a range of \feh\ values, centered upon the
preferred value for each of the clusters.  Most of the models used a
scaled solar heavy element composition \citep{gre93}. The models for 47 Tuc
were calculated with $[\alpha/{\rm Fe}] = 0.0$ (scaled solar) and
$[\alpha/{\rm Fe}] = +0.40$.  

The calibrated solar model for this study had a helium mass fraction
of $Y_\sun = 0.263$ and a heavy element mass fraction of $Z_\sun =
0.0182$.  The primordial helium abundance was taken to be $Y_p =
0.234$ \citep{olive}, and the helium abundance for each model was
calculated using the relation $Y = Y_p + (\Delta Y / \Delta Z) * Z$
with $\Delta Y / \Delta Z = (Y_\sun - Y_p)/ \! Z_\sun$.

\begin{figure}
\centerline{\epsfig{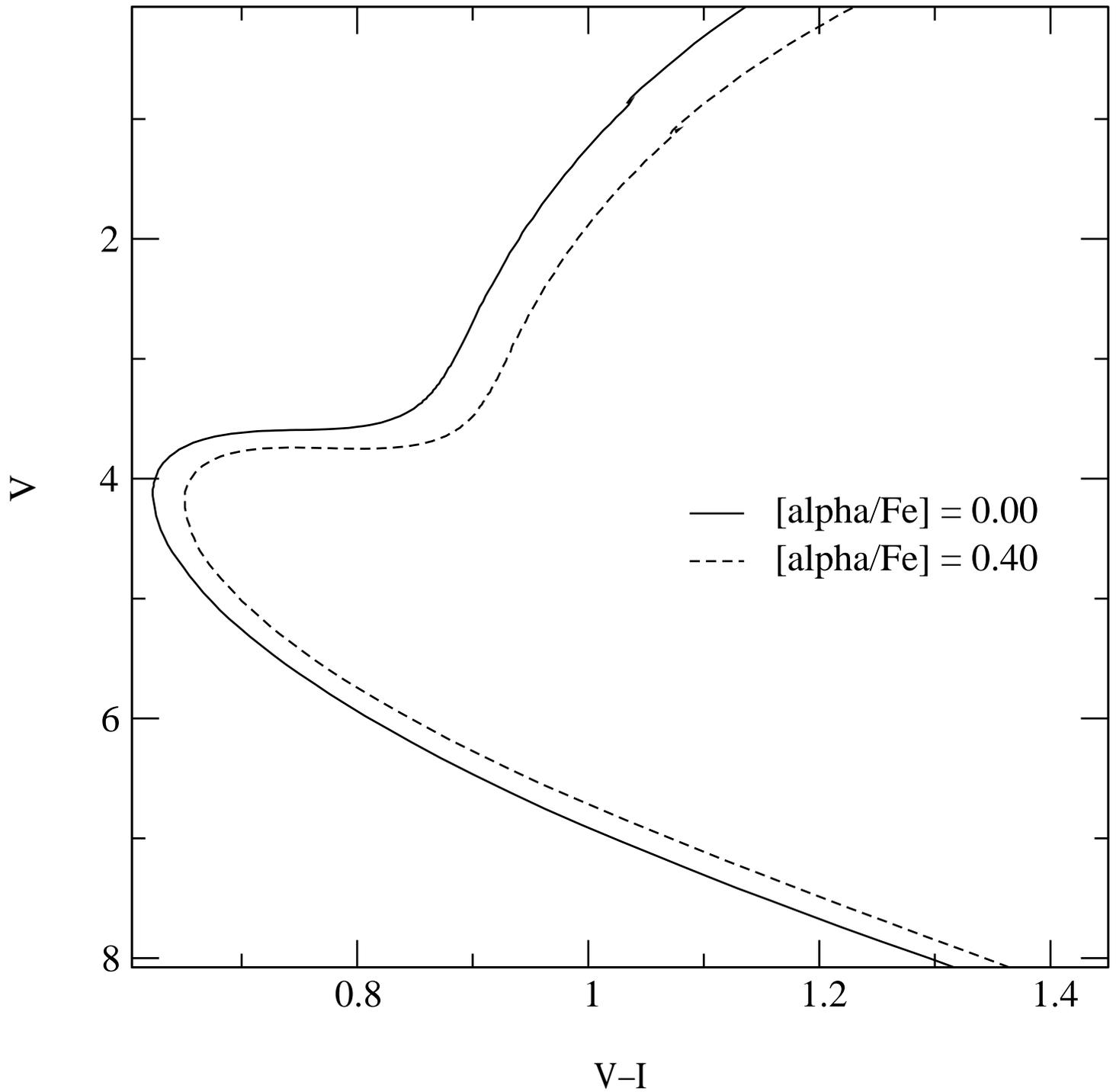}}
\caption{The 10 Gyr isochrones with $\feh = -0.70$ for both
the $\alpha$-enhanced (0.40 dex) and the non $\alpha$-enchanced cases.  The shift
in the isochrone corresponds to a change in $\feh$ of about 0.20 dex.}
\label{figalpha}
\end{figure}

The isochrones were constructed for ages between 5 and 14 Gyr, in 1
Gyr increments for each of the compositions.  Color transformations to
the observational plane were done in a manner identical to that
described in CGL.
Figure \ref{figalpha} compares the 10 Gyr isochrone for
$\feh = -0.70$ for both the $\alpha$-enhanced and the non
$\alpha$-enhanced cases.  The resulting change in the isochrone
corresponds to a change in $\feh$ of nearly 0.20 dex.  These results
are similar to those found in \citet{Sal98}.  

\begin{figure}
\centerline{\epsfig{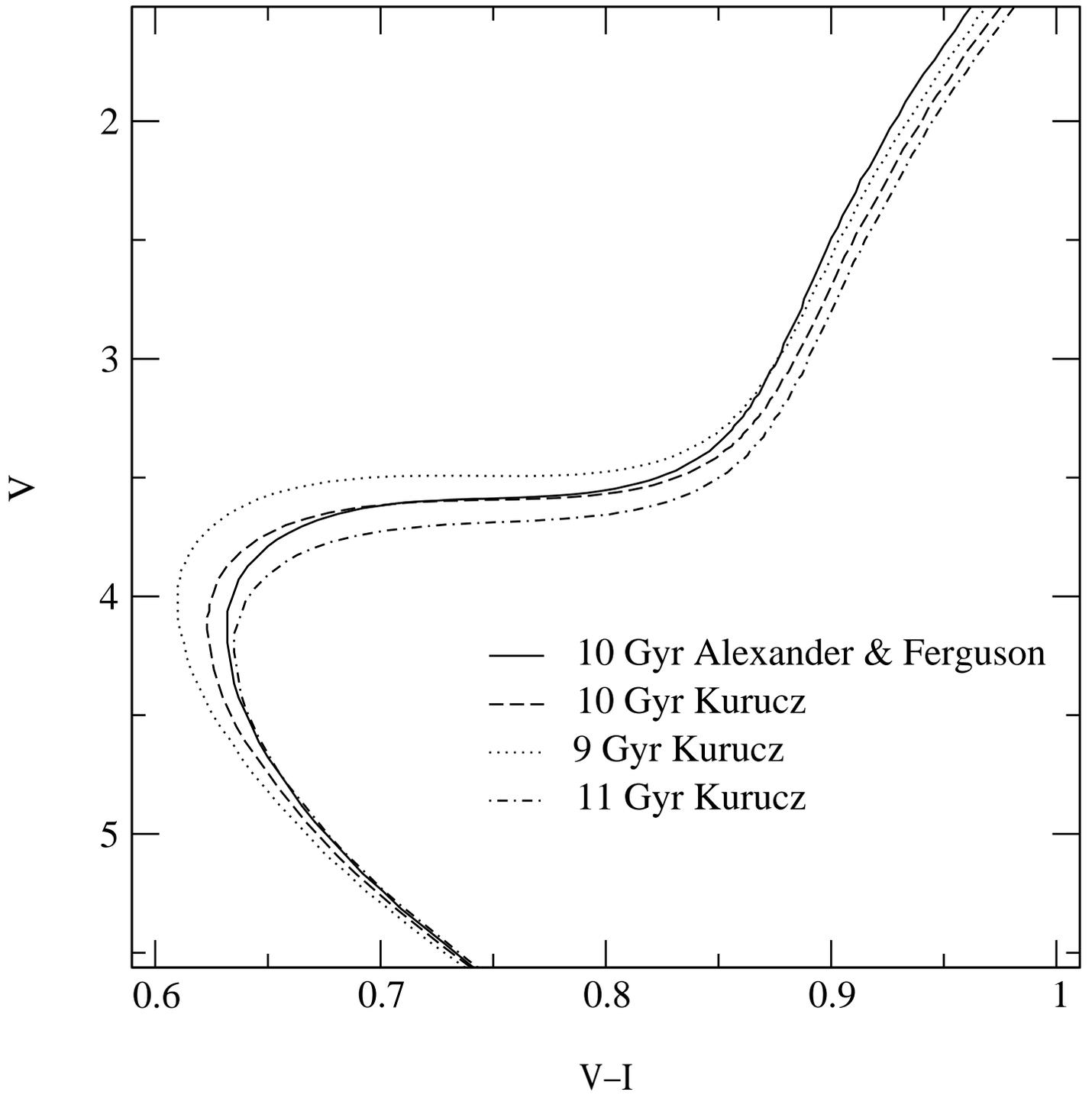}}
\caption{The 10 Gyr isochrone generated using Alexander \& Ferguson opacities, with $\feh = -0.70$.
Also shown are the 9, 10, and 11 Gyr isochrones using Kurucz opacities. }
\label{figalex}
\end{figure}

Additional isochrones were generated in order to ascertain the effect
that changes in the low temperature opacities would have on the
isochrones.  Figure \ref{figalex} plots the 10 Gyr isochrone for $\feh
= -0.70$ with \citet{Alex94} opacities, as well as the 9, 10 and 11
Gyr isochrones with \citet{Kur91} opacities used in this study.  This
figure demonstrates that ages determined using isochrones generated
with Kurucz opacities will indicate ages about 0.5 Gyr older than
isochrones using Alexander \& Ferguson opacities.  This analysis was
done in order to compare the age of NGC 6791 as found in CGL, which
used Alexander \& Ferguson opacities in its models, with the ages of
the clusters in this study and CSA, which utilized Kurucz opacities.
Kurucz (1991) opacities were used in this study and by CSA as these
opacities are available for both scaled solar and $\alpha$-element
enhanced compositions.  The Alexander \& Ferguson opacities were only
available to us for scaled solar compositions.

\section{Thin Disk Ages \label{thindisk}}
\subsection{NGC 188}
\begin{figure}
\centerline{\epsfig{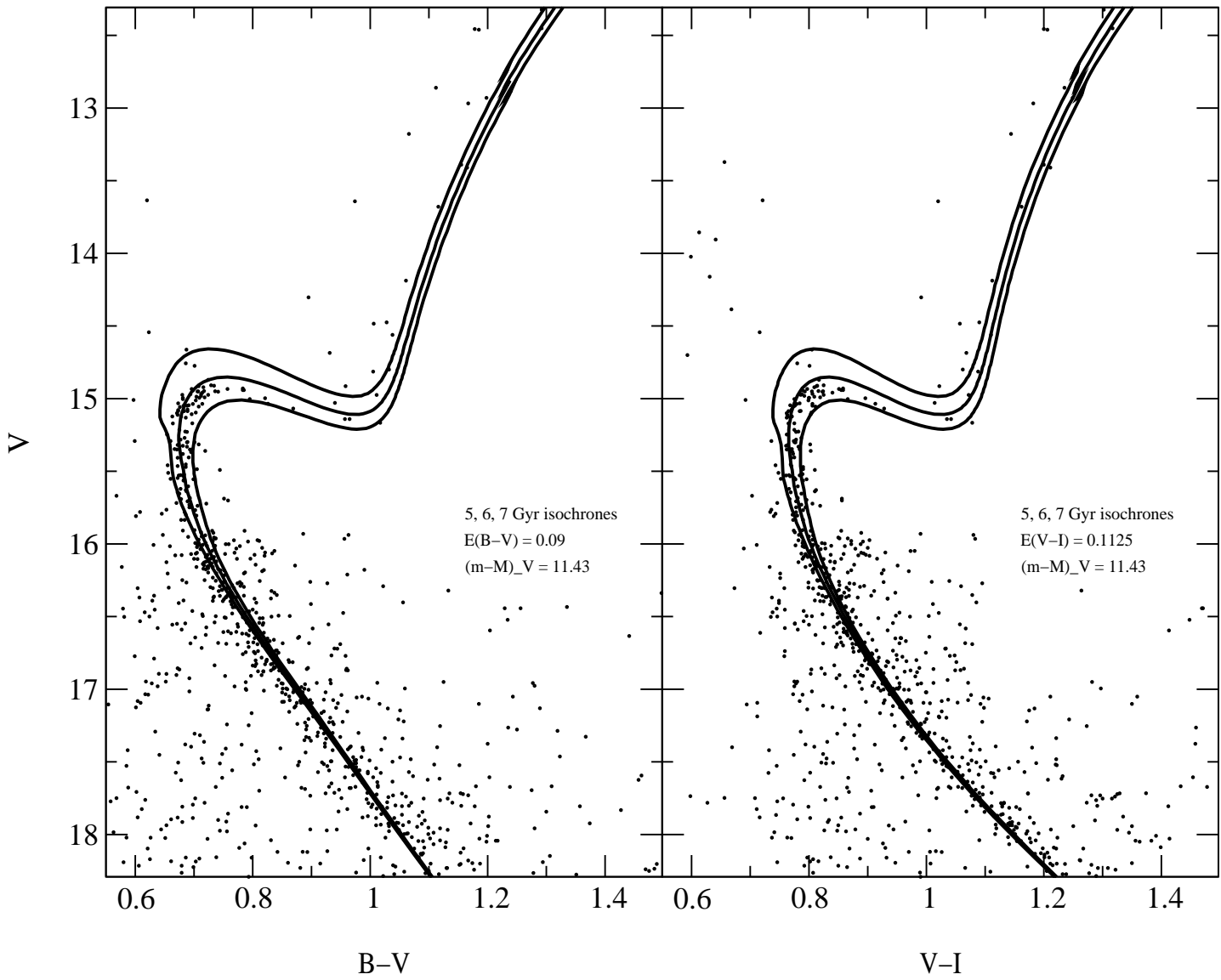}}
\caption{The 5, 6, and 7 Gyr isochrones with $\feh = -0.05$ fit simultaneously in B-V and V-I 
to the CMD for NGC 188.  The photometric data was taken from \citet{Sar99}.  }
\label{fign188}
\end{figure}

Isochrones were fit to the CMD for NGC 188 obtained from observations
by \citet[ hereafter S99]{Sar99}.  The CMD was cross-correlated to
include only cluster members with a greater than 75 percent membership
probability \citep{Dinpriv}.  Membership criteria was based upon a
proper motion study by \citet{Din96}.  
NGC 188 has $\feh = -0.05$ based upon several determinations
\citep{Twar97,Thog93,Hob90}.  Isochrones with $\feh = -0.05$ were fit
simultaneously in \bv\ and \vi\  using the main sequence, allowing the
distance modulus to vary around the value established by S99.  Figure
\ref{fign188} shows the 5, 6, and 7 Gyr isochrones 
fit to the observational data.  The reddening was taken to be $\ebv =
0.09$ (S99) and transformed to $\evi$ by $\evi = 1.25*\ebv = 0.1125$.
The distance modulus which yielded an excellent main sequence fit was
$\dmv = 11.43$, well within the error bars for the value determined by
S99 of $\dmv = 11.44\pm0.08$.  The age indicated by both \bv\ and \vi\
fits was 6.5 Gyr.

\begin{figure}
\centerline{\epsfig{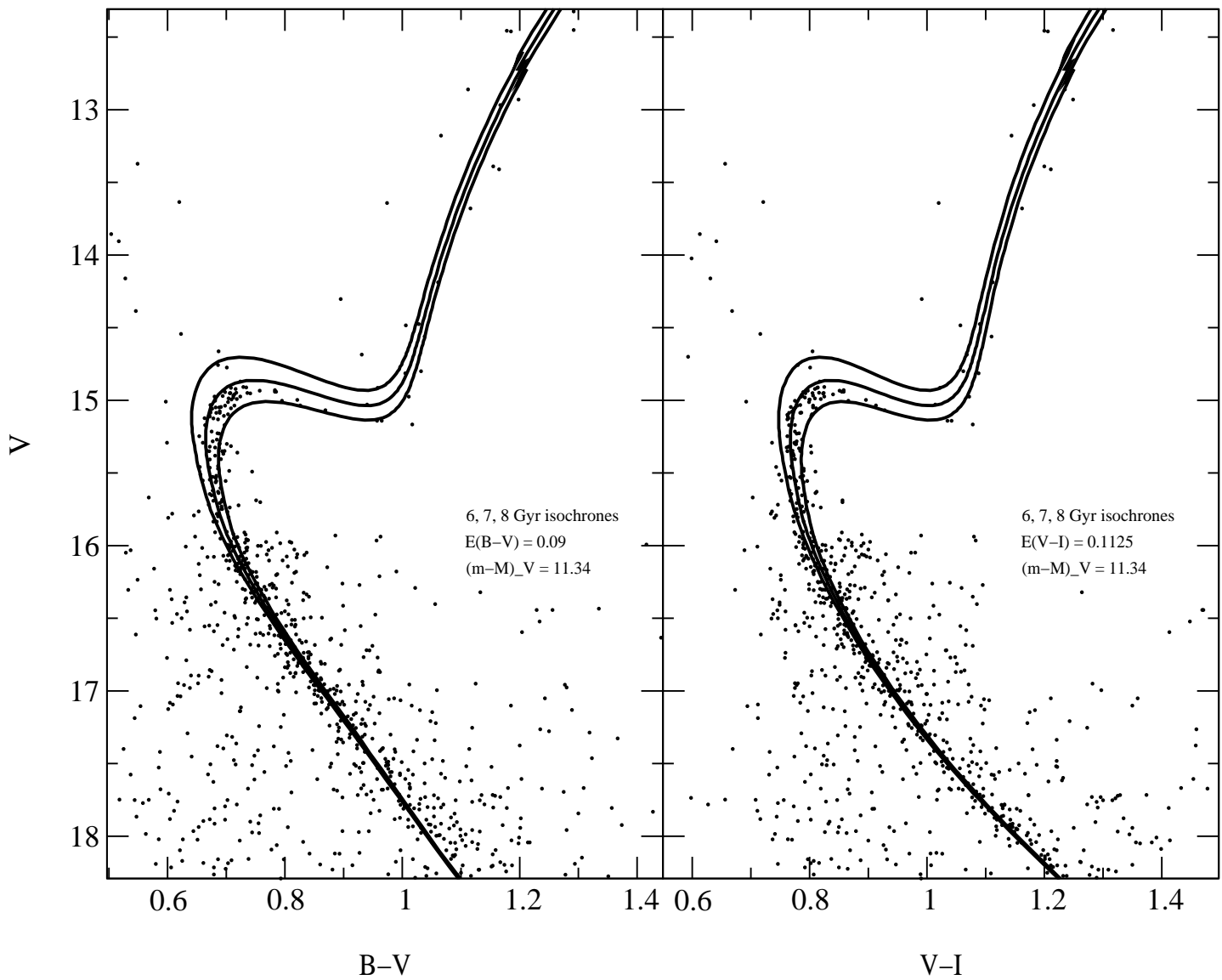}}
\caption{The 6, 7, and 8 Gyr isochrones with metallicity of $\feh = -0.15$ 
fit to NGC 188.  }
\label{fign188lml}
\end{figure}

\begin{figure}
\centerline{\epsfig{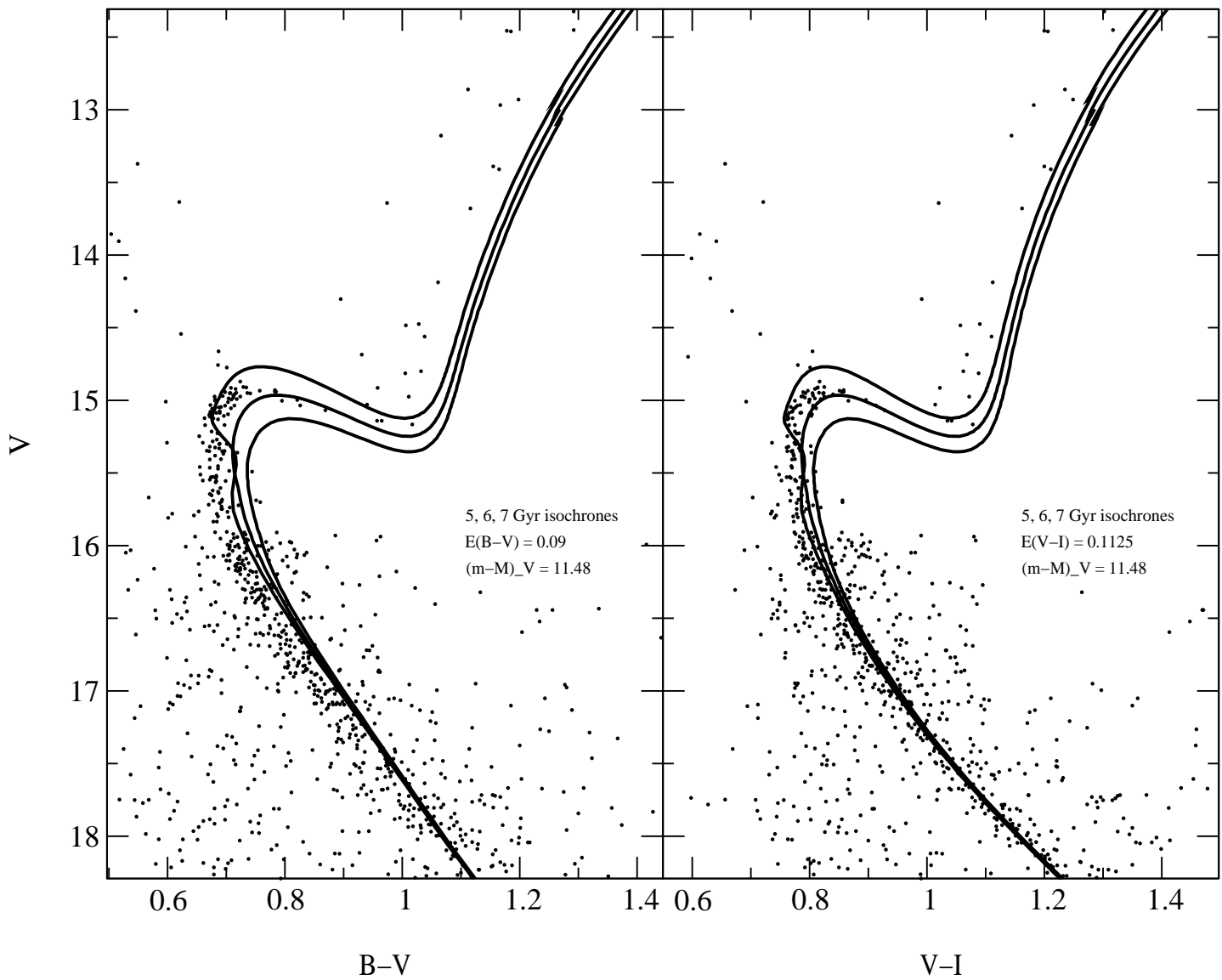}}
\caption{The 5-7 Gyr isochrones with metallicity of $\feh = +0.05$
fit to NGC 188.  }
\label{fign188uml}
\end{figure}

Isochrones for a more metal poor composition of
$\feh = -0.15$ were fit to the CMD, again assuming a reddening of
$\ebv = 0.09$.  The fits produced a distance modulus of $\dmv =
13.34$, slightly less than S99, and an age of 7.5 Gyr.  Figure
\ref{fign188lml} shows the 6, 7, and 8 Gyr isochrones for this lower
metallicity fit to the observational data.  Isochrones with a higher
metallicity, $\feh = +0.05$, were fit to the data as well and are
shown in Figure \ref{fign188uml}, indicating $\dmv = 11.48$ and an age
of 5.5 Gyr.  However, it would appear unlikely that a younger age as a
result of higher metallicity is the case, as the CMD does not show a
hook at the main sequence turn off which is seen in the isochrones.

In order to determine the uncertainty due to reddening, the $\feh =
-0.05$ isochrones were fit for $\ebv = 0.07$ and $0.11$.  These fits
indicate ages of 7.5 Gyr and 5.5 Gyr, respectively.  The three sets of
isochrone fits for NGC 188 are summarized in Table \ref{tabn188}.  The
best age estimate was found to be $6.5\pm1.0$ Gyr, allowing for
uncertainty in the metallicity and reddening of the cluster.  This
error estimate does not include the uncertainty in the input physics
in the stellar models and isochrones, and so should be viewed as the
error on the relative age of NGC 188.

\subsection{Berkeley 17}
Berkeley 17 (Be 17) has been suggested to be the oldest open cluster
\citep{phelps}.  It has a metallicity of $\feh = -0.29\pm 0.13$ from
moderate resolution spectroscopy \citep{friel}, while \citet{carr2}
estimate $\feh \sim -0.35$ based upon the slope of the red giant
branch in the infrared\footnote{We note that the JK photometry presented by
\citet{carr2} contains a great deal of scatter on the main sequence
and so we are unable to use this data in our isochrone fitting
procedure.}.

\begin{figure}
\centerline{\epsfig{file=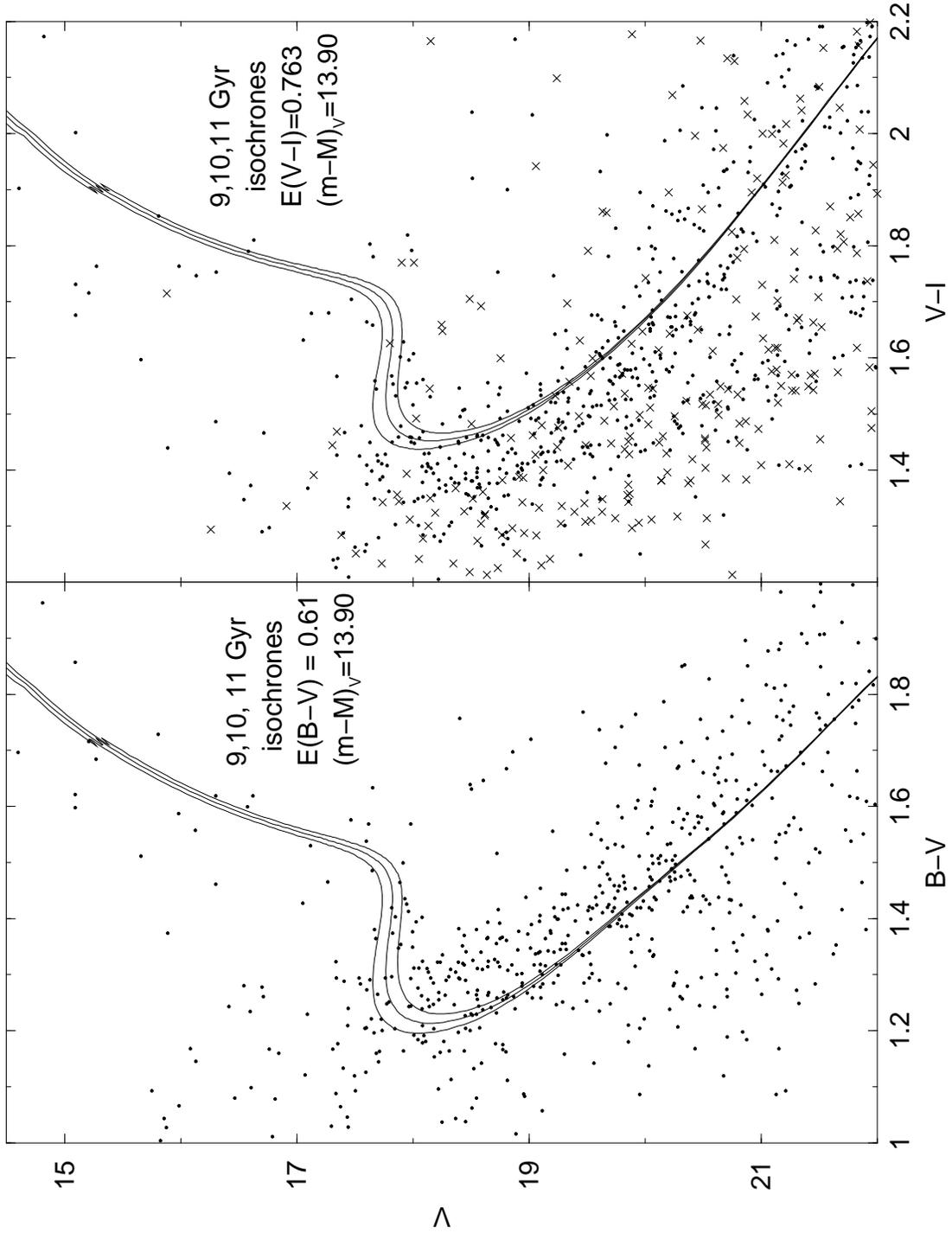,height=19.0cm}}
\caption{The best fit of our isochrones to Be 17. The Be 17 data is
shown with filled circles, while the field stars (from a nearby field)
are shown with Xs in the \vi graph.  This is clearly not
a good fit, and as a result we are unable to assign an age to Be 17.}
\label{figbe17}
\end{figure}

\citet{phelps} obtained BVI photometry of
this cluster which was used in our isochrone fits.  A simultaneous fit
to the \bv\ and \vi\ photometry was attempted, assuming 
$\evi = 1.25*\ebv$.  
We attempted to fit isochrones with $\feh = -0.10,
-0.24,$ and $-0.40$ with $[\alpha/{\rm Fe}] = +0.0$ and $+0.40$ (a
total of 6 different compositions) without success.  In
all cases, the Be 17 main sequence was redder than our isochrones in
\bv\ and bluer than our isochrones in \vi.  Our best attempt at a fit
is shown in Figure \ref{figbe17}.  As this is clearly not an
acceptable fit to the data, we are unable to assign an age to Be 17.  
There are a number of possible explanations for the inability of the
isochrones to simultaneously fit the \bv\ and \vi\ data.  These include:
(1) a non-standard extinction law in the direction of Be 17
(ie.\ $\evi \neq  1.25*\ebv$), 
(2) a  helium abundance significantly different than in our
models, 
(3) errors in our isochrones 
(4) errors in the photometry (of order 0.05 mag in the color).

\subsection{Hipparcos Field Stars \label{fieldstars}}
Open clusters are expected to dissipate/disrupt in the galactic plane.
Thus, only the most massive open clusters or those with orbits which
keep them far away from the plane for most of their lifetimes are
expected to survive for signficant periods of time.  For this reason,
the age of the oldest field stars in the thin disk may provide a
better estimate for the age of the thin disk than open clusters.  An
analysis of the age of oldest thin disk stars in the solar
neighborhood was made by comparing solar metallicity isochrones to a
selected sample of field stars from the Hipparcos database
\citep{hipp}.  

The Hipparcos database was searched for single, non-variable stars
with good parallaxes ($\sigma_\pi/\pi < 0.12$).  These stars were
cross referenced with the 1996 \feh\ catalogue \citep{fehcat} to find
stars with $\feh$ within 0.07 dex of solar.  Choosing stars with
metallicities near the solar value should ensure that these stars are
members of the thin disk.  In addition, choosing a relatively
narrow metallicity range makes the age determination process simple.
The selected stars were then plotted on a CMD and
compared to solar metallicity isochrones.  As many of the stars do not
have I band photometry, the comparison to the isochrones was only done
in \bv.

\begin{figure}
\centerline{\epsfig{file=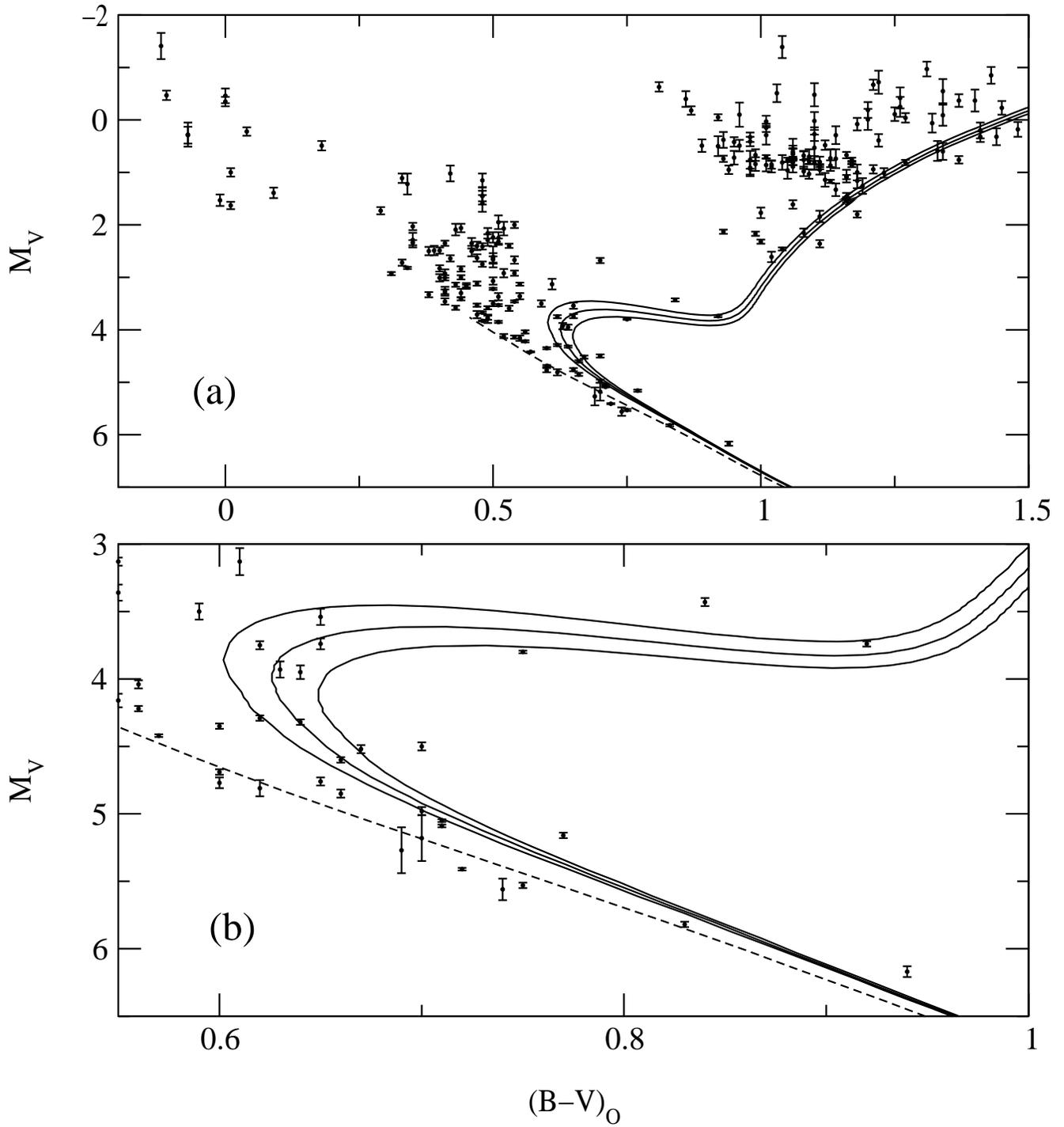,height=19.0cm}}
\caption{Top: Hipparcos field stars overlaid with the 6-8 Gyr solar metallicity
isochrones.  Bottom: An enlargement of the turnoff region. }
\label{fighipp}
\end{figure}

We are interested in determining the onset of star formation in the
thin disk, and so concentrated our attention on the oldest stars in
our sample for which it is possible to determine ages.  These are the
faintest stars in the main sequence turn-off or subgiant branch
regions of the CMD.  There were 21 stars which
appeared to be old in our search of the Hipparcos database.  These 21
stars were checked for other published metallicities and space
velocities.  Some stars were omitted from the sample which were shown
in other sources to have metallicities of more than 0.07 dex from
solar \citep{Edv93,McW90} or shown to have space velocities which are
not characteristic of the thin disk \citep{Egg98,Edv93}.  The
resulting sample of stars are shown in Figure \ref{fighipp} along with
our zero-age main sequence and 6 -- 8 Gyr isochrones (all with $\feh =
0.0$).

Figure  \ref{fighipp}b reveals that there are
3 stars which are much brighter/redder than our main sequence isochrones.
There a few possible reasons for this discrepency, including that
these stars are binaries, or are more metal-rich than indicated by the 
1996 \feh\ catalogue.  We will not determine the age of these three
stars.  

An inspection of the oldest stars in the main sequence turn-off and
subgiant branch region in Figure \ref{fighipp}b reveals that 2 stars
lie on the 8 Gyr isochrone, 4 stars lie on or very near the 7 Gyr
isochrone and 1 star lies between the 7 and 8 Gyr isochrone. The
derived ages for the stars in the turn-off region are quite sensitive
to possible errors in the colors of the stars.  For example, if the
\bv\ color of the 8 Gyr star with $M_V = 4.53$ was changed from and
$\bv = 0.670$ to $\bv = 0.655$ then its age would be changed to 7 Gyr.
Given the uncertainties in the photometry, reddening, and in our color
transformation we are reluctant to base our estimate for the age of
the oldest stars in the thin disk on just two stars.  For this reason,
we estimate that the oldest stars are $7.5\pm 0.5$ Gyr old.
Isochrones for $\feh = -0.05$ and +0.05 yielded ages of 8.0 Gyr and
7.0 Gyr respectively and we conclude that the oldest solar
metallicity stars in the thin disk have an age of $7.5\pm 0.7$ Gyr.
In determining this error, we simply added in quadrate the $\pm 0.5$
Gyr error due to the uncertainty in the metallicity of the stars with
the $\pm 0.5$ Gyr uncertainty due to the photometry/reddening and
color transformation.

There is a range in metallicities in the thin disk and \citet{Edv93}
found that (in the local solar neighborhood) the thin disk has
metallicities ranging down to $\feh \simeq -0.20$.  For this reason,
we elected to search the Hipparcos database for single, non-variable
stars with good parallaxes ($\sigma_\pi/\pi < 0.12$) with $-0.08 \leq
\feh \leq -0.25$ (\feh\ values from the 1996 \feh\ catalogue
\citet{fehcat}).  This sample of stars allows us to determine the age
of the somewhat  metal-poor stars in the thin disk.  In order to
focus on the oldest stars, color and magnitude cuts were made in order
to select stars with $3.4 < {\rm M_V} < 5.0$ and $ 0.54 < \bv <
0.74$.  These values were chosen based upon an inspection of our 7 Gyr
and older isochrones with $\feh = -0.10$ and $\feh = -0.24$.  These
cuts were made rather generous to ensure that all potentially older
stars are included in the sample.  These cuts resulted in a 
sample of 21 stars which were plotted on a CMD
with isochrones of appropriate metallicities.  Stars which
were less than $\approx 5$ Gyr old were deleted from further study.
This left a sample of 16 stars. 

Basic data for each of these 16 stars was retrieved from the SIMBAD
database at CDS.  Four stars which were classified as variables or
spectrospic binaries were removed from list as it would not be
possible to determine their ages.  The derived ages are quite
sensitive to the metallicity of the star.  For this reason, the \feh\
references were checked for each star and 3 stars whose recent (post
1995) \feh\ determinations were significantly different than those
listed in the 1996 \feh\ catalogue were removed from the list.  
The basic data for each
of these stars are given in Table \ref{tabfield}.  The parallaxes,
proper motions and colors are from the Hippacos catalog.  The \feh\
values are from \citet{fehcat}, and the radial velocities were
obtained from the SIMBAD database.  

The age of each of these nine stars was determined from our isochrones
(interpolating between the $\feh = -0.10$ and $\feh = -0.24$
isochrones).  The error in the derived age of each star is due to the
error in the absolute magnitude (due to the error in the parallax), an
assumed color error of $\pm 0.005$ in \bv\ and an assumed error in
\feh\ of $\pm 0.05$ dex.  The ages of these stars are given in Table
\ref{tabfield2}, which also includes some derived kinematic data and
orbital parameters for each star.  The kinematic data and orbital
parameters allow us to separate thin and thick disk stars.

The space motion of each star was calculated using a routine kindly
provided to us by Dr.\ Dana Dinescu.  The motions were calculated in a
cylindrical coordinate system (origin at the Galactic center) by
adopting a solar radius of $R_\sun = 8.0$ kpc and a rotation velocity
of the local standard of rest (LSR) of $\Theta_0 = 220$ km/s.  In this
cooridinate system, the $\Pi$ component is positive outward from the
Galactic center, $\Theta$ is positive in the direction of Galactic
rotation and $W$ is positive towards the north Galactic pole.  Errors
in the derived velocities include errors in the proper motions, radial
velocities and distances.

Table \ref{tabfield2} includes some basic orbital parameters
based upon intergation of the orbits in two models of the Galaxy's
potential.  These integrations were made by Dr.\  Dana Dinescu, and
full details of the integration routine may be found in \citet{dana}.
The orbits were integrated in Galactic potential models given by
\citet{jsh95} and \citet{pac90}.  The two potentials yield similar
orbital parameters and only the orbital parameters from the
\citet{pac90} potential are shown in Table \ref{tabfield2}.  The
orbital parameters listed in Table \ref{tabfield2} are:
$L_z$ the $z$-component of the angular momentum (a conserved
quantity), pericentric ($R_{per}$) and apocentric ($R_{apo}$) radii,
the maximum distance from the plane $z_{max}$, the eccentricity of the
orbits $e$ and the inclination angle with respet to the Galactic plane
$\Psi$.  

The stars in Table \ref{tabfield2} all have $-0.14\le \feh \le -0.25$ and in
this metallicity range one finds both thin and thick disk stars in the
local solar neighborhood.  The dispersion in the kinematics
of the thin and thick disk make it impossible to definitavily assign a
single star to either the thin or thick disk.  For example, in the
thin disk $\sigma(W) = 20$ km/s while in the thick disk $\sigma(W) =
40$ km/s \citep{Edv93}.  Thus, the $W$ velocity of HD 41330 ($W = -25.1$
km/s) implies that it could be member of the thick or thin disk.  By
considering all of the kinematic and orbital parameters ($\Pi$,
$\Theta$, $W$, $L_z$, $e$ and $\Psi$) one can classify a star as
either a thin or thick disk star, bearing in mind that these
classifications will never be 100\% accurate.  In general, thin disk
stars have small $\Pi$ and $W$ velocities, $\Theta \approx 220$
km/s,  $L_z \approx 1700$ kpc km/s, $z_{max} \la 0.3$ kpc, $e \la 0.2$
and small values of $\Psi$.  

In Table
\ref{tabfield2} we have indicated the most propobable classificiation
of each star (thin or thick disk) {\em based on the kinematic and
orbital parameters alone}.  HD 52711 and 210918 are prototypical
thick disk stars with rotation ($\Theta$) velocities significantly
different than the LSR, a low $z$-component to their orbital angular
momentum ($L_z$) and a high eccentricity.   In contrast, HD 207129 is a
prototypical thin disk star with a rotation velocity and $L_z$ similar
to the LSR, a low eccentricity, a small $z_{max}$ and a small angle to
the Galactic plane.  

Three stars which are likely thin disk stars (HD 15335, 202628 and
207129) have similar ages of $\simeq 6.4$ Gyr.  The oldest stars which
appear to be thin disk stars are HD 32923 ($10.0\pm 0.5$ Gyr) and HD
41330 ($9.3\pm 0.6$ Gyr).  Both of these stars have rotation
velocities and $L_z$ similar to the LSR and low eccentricities.  The
$z_{max}$ of these stars is not too large (0.46 and 0.29 kpc) given
that thin disk stars have $\ge 0.325$ kpc scaleheight exponentials
\citep{maj}. The angle of their orbits to the Galactic plane ($\Psi =
3.45^{\degr}$ and $2.14^{\degr}$) are somewhat larger than typical for
thin disk stars but not extremely so. As a result, we believe that it
is likely that at least one of these two stars is a true member of the
thin disk.  These stars are located in a region of the CMD where the
derived ages are relatively insensitive to the metallicities, colors
and absolute magnitudes.  Consequently,the derived ages have very
small error bars.  Averaging the ages of these two stars together we
find that the oldest, somewhat metal-poor thin disk stars in the solar
neighborhood have an age of $9.7\pm 0.6$ Gyr.

The two stars with kinematic and orbital parameters most representive
of the thick disk (HD 52711 and 210918) have quite different ages of 
$6.7\pm 1.0$ Gyr and $11.7\pm 1.1$ Gyr.  The older age is similar to
47 Tuc (see below) while the younger age would suggest that there is a
considerable overlap in the ages of the thin and thick disks. However,
we are reluctant to reach such a conclusion based only on one star.
Two other stars which might be thick disk stars (HD 11007 and 67458)
are also fairly young ($6.6\pm 0.5$ and $6.5\pm 1.0$).  However, one
could argue that the kinematics and orbital parameters of these two
stars are not too different from the thin disk, and so there
identification as thick disk stars is debateable.  Further age
determinations of more metal-poor stars ($\feh < -0.25$)  are needed
before one can conclude that their is a significant spread in the age
of the thick disk.


\section{47 Tuc\label{sec47tuc}}
\subsection{Isochrone fitting ages}
Isochrone fitting ages were determined for the thick disk globular
cluster 47 Tuc.  The photometric data for the cluster was obtained
from \citet{Kal98}.  Heavy element abundances in the literature
indicate values of $\feh = -0.70\pm0.07$ \citep{Crr97}, and $\feh =
-0.81$ \citep{Bro92}.  Additionally, \citet{Bro92} indicate an
enhancement in $\alpha$-capture elements of $\afe = 0.22$.  However,
for the stellar models used in this study, opacities were only
available for $\afe = 0.00$ and $0.40$.  Thus isochrones were fit for
abundances of $\feh = -0.70$ and $-0.80$ both with $\afe = 0.40$.

\begin{figure}
\centerline{\epsfig{file=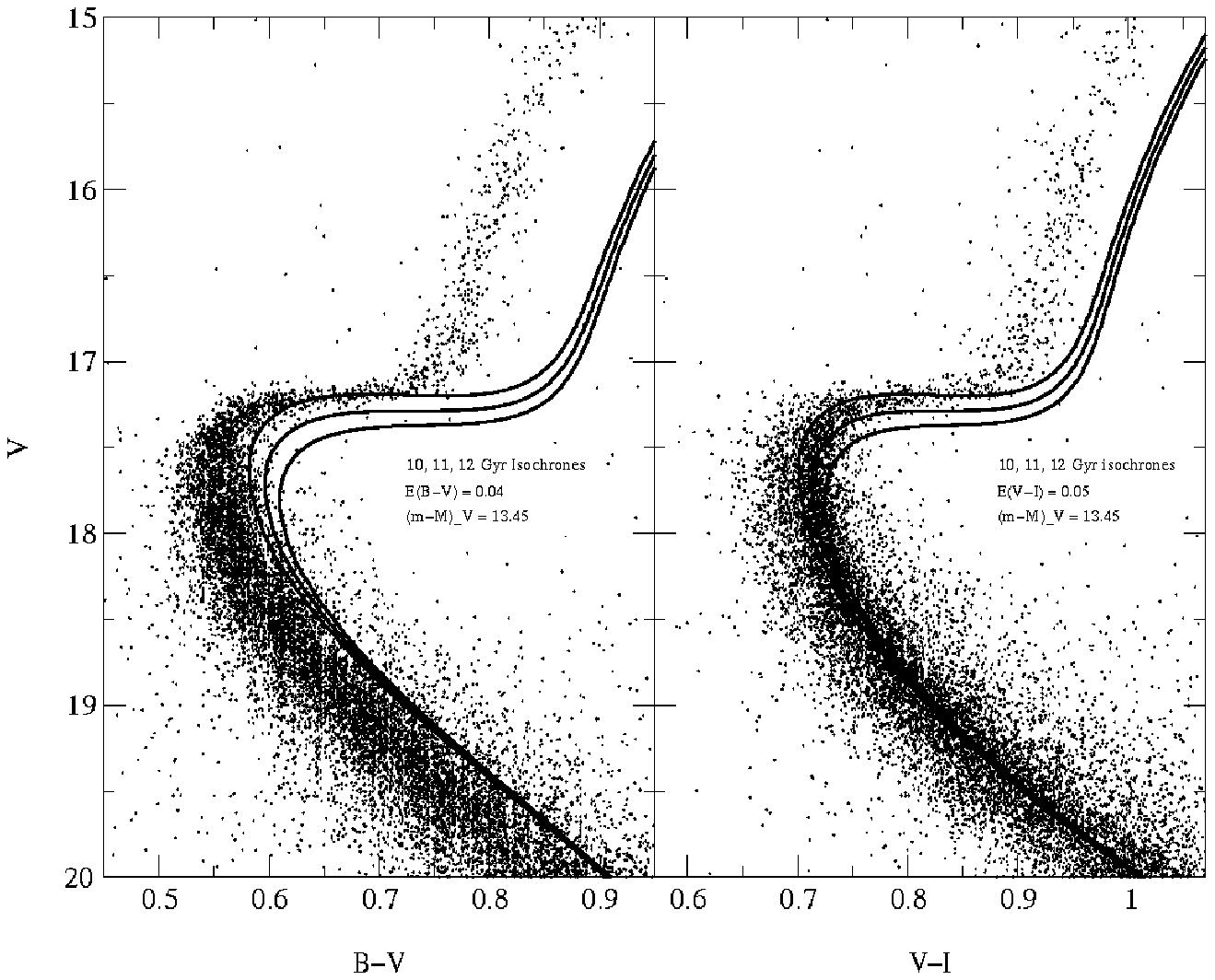,height=15.0cm,angle=90}}
\caption{The 10, 11, and 12 Gyr isochrones with $\feh = -0.80$ and
$[\alpha/{\rm Fe}] = +0.40$ fit to the CMD for 47 Tuc.  }
\label{fig47tuc-080}
\end{figure}

\begin{figure}
\centerline{\epsfig{file=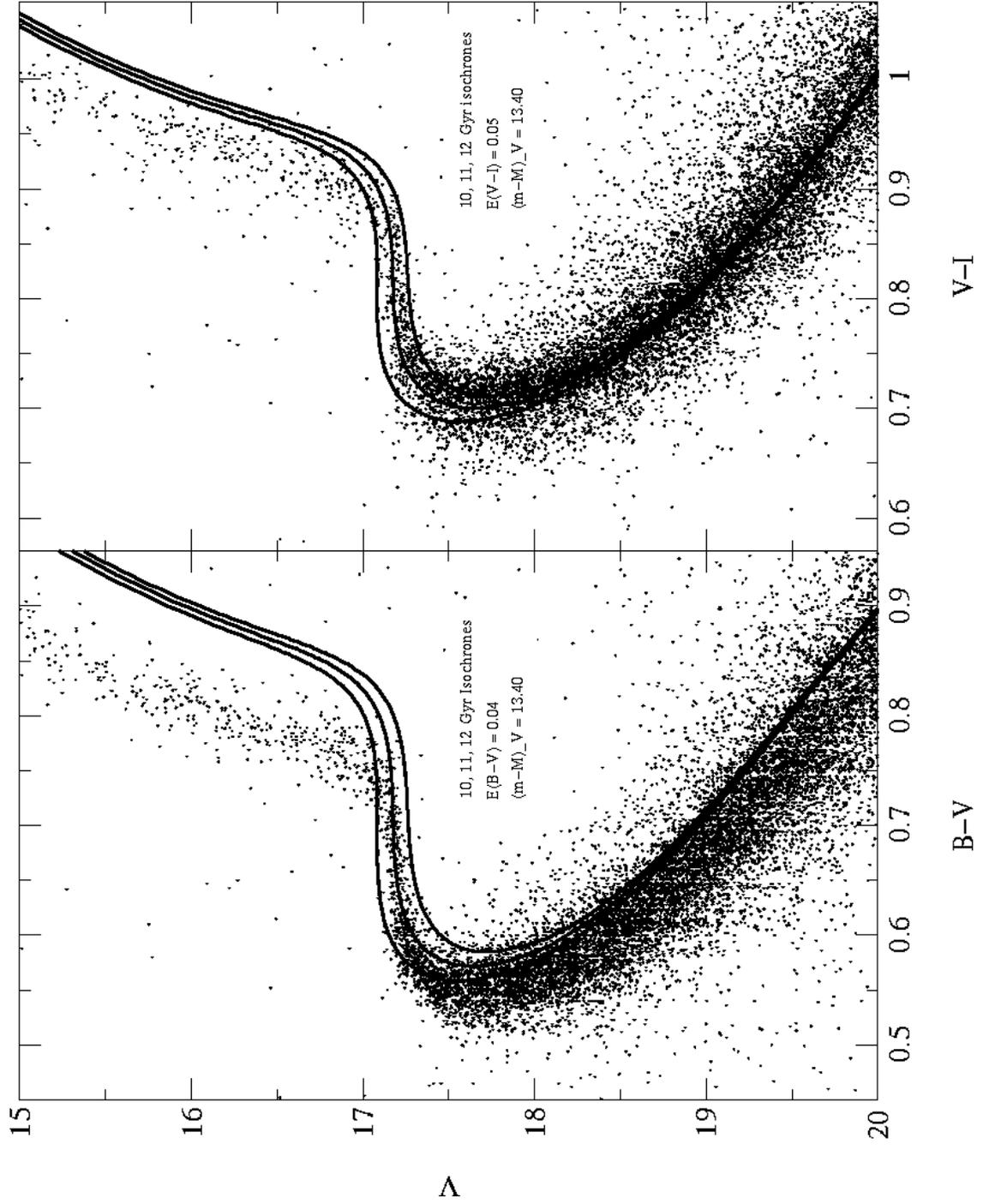,height=15.0cm,angle=90}}
\caption{The 10, 11, and 12 Gyr isochrones with $\feh = -0.70$ and
$[\alpha/{\rm Fe}] = +0.40$ fit to the CMD for 47 Tuc.  }
\label{fig47tuc-070}
\end{figure}

\begin{figure}
\centerline{\epsfig{file=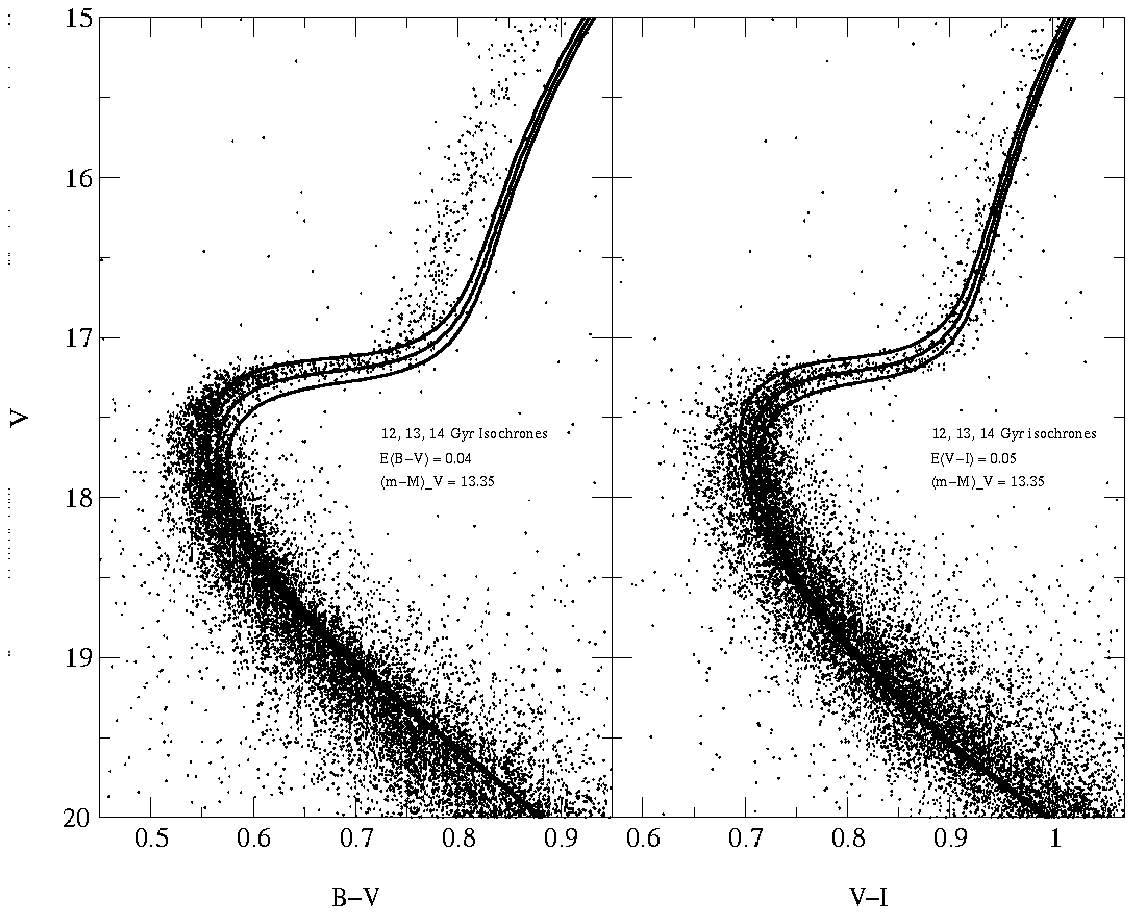,height=15.0cm,angle=90}}
\caption{The 12, 13, and 14 Gyr isochrones with $\feh = -0.95$ and
$[\alpha/{\rm Fe}] = +0.40$ fit to the CMD for 47 Tuc.  }
\label{fig47tuc-095}
\end{figure}

Isochrones were fit in a manner identical to NGC 188, simutaneously in
B-V and V-I.  The value of reddening was fixed at $\ebv = 0.04$
\citep{Har96} and $\evi = 1.25*\ebv = 0.05$.  The distance modulus was
varied around $\dmv = 11.37$ \citep{Har96} in order to obtain a good
main sequence fit.  Figure \ref{fig47tuc-070} shows the isochrone fits
for $\feh = -0.70$ and $\afe = 0.40$.  The poor simultaneous fit is
likely due to the difference in $\alpha$-enhancement by nearly 0.2 dex
between the literature value and the models.  The fits for $\feh =
-0.80$ and $\afe = 0.40$, shown in Figure \ref{fig47tuc-080}, likewise
indicate a poor simultaneous fit to B-V and V-I.  In order to better
reflect the total heavy element abundance ($Z$) in the cluster,
isochrones were generated with $\feh = -0.95$ and $\afe = 0.40$.  This
abundance was chosen to match the heavy element mass fraction for the
values of \feh\ and \afe\ found in \citet{Bro92}, corresponding to $Z
= 0.0040$.  The result, shown in Figure \ref{fig47tuc-095}, was an
acceptable fit which indicated an age of $12.5\pm 0.5$ Gyr for a
reddening of $\ebv = 0.04$ and distance modulus of $\dmv =
11.35$. Both the reddening and distance modulus were in good agreement
with the values quoted in \citet{Har96}.  Uncertainty due to reddening
was determined by producing fits with $\ebv = 0.03$ and $0.05$.  These
fits indicated ages of $14\pm 0.5$ and $11.5\pm 0.5$ Gyr,
respectively.  Table \ref{tab47tuc} summarizes the isochrone sets and
parameters used in fitting to 47 Tuc.  The age used for comparison to
the other clusters and field stars was $12.5\pm1.5$ Gyr, as 
indicated and includes the uncertainty due to reddening.

\subsection{$\dvbto$ ages}
Age determinations were also made using the $\dvbto$ method described
in CSA.  The $\dvbto$ determination method is theoretically robust,
but requires a well defined horizontal branch not present in the CMDs
of open clusters.  In order to compare the $\dvbto$ ages in CSA with
the isochrone fitted ages in this paper, the age of 47 Tuc was
determined using both methods, and all ages were calculated relative
to 47 Tuc.  The observed value for $\dvbto$ in the CMD for 47 Tuc was
3.18 $\pm$0.04 mag.  The theoretical value of $V_{\rm HB}$ was
calculated  using $\mvrr = 0.23 * (\feh + 1.6) + 0.56$ (CSA) and
corrected for the fact that the HB was only apparent redward of the RR
Lyrae instability strip in 47 Tuc.  This correction was determined
using the theoretical HB models of \citet{Dem00}.  Using the $\feh =
-0.95$, $[\alpha/{\rm Fe}] = +0.40 $ isochrones yielded an age of $13.5$ Gyr.
The $\feh = -0.70$, $[\alpha/{\rm Fe}] = +0.40$ isochrones yielded an
age of 12.2 Gyr.  47 Tuc has $[\alpha/{\rm Fe}] = +0.22$ \citep{Bro92} and
$\feh = -0.70\pm0.07$ \citep{Crr97} which gives a $Z$ value
intermediate between the above 2 isochrones sets.  For this reason, we
average the 2 ages and adopt an \dvbto\ age for 47 Tuc of $12.9\pm
0.7$ Gyr.

The age determined from the \dvbto\ is $0.4$ Gyr older than that
determined from the isochrone fits. This small difference in age is
well within the estimated error in our isochrone fit age.  As the
\dvbto\ age estimate is expected to yield more accurate absolute ages
than isochrone fitting, we will assume an absolute age of $12.9$ Gyr
for 47 Tuc.  However, when determining the relative age of 47 Tuc to
the old thin disk, we will use the isochrone fitting age of 47 Tuc, as
the open clusters and Hipparcos field stars had their ages determined
using isochrone fitting.


\section{Summary and Discussion \label{secsumm}}
The age of 47 Tuc ($12.9\pm 0.7$) Gyr provides a reliable estimate for
the age of the thick disk.  The results for the various relative
stellar ages determined in this study are summarized in Table
\ref{tabresults}. This table includes the age for NGC 6791 from CGL
and the ages of three inner halo globular clusters (NGC 6652, NGC 1851
and M107) investigated by CSA.  The results from CSA indicate that
star formation in the inner halo began roughly at the same time as the
thick disk.  All three of the studies discussed here (CSA, CGL, and
this study) used the same input physics and distance scale, allowing
for a relative age comparison between all clusters in the studies.
The relative age between the thin disk stars and 47 Tuc in this study
and CGL were determined using the isochrone fitting ages.  Relative
ages between 47 Tuc and the inner halo were determined using the
$\dvbto$ method.  The globular cluster 47 Tuc was used as a ``bridge''
between the studies, as its age was determined using both the
isochrone fitting and $\dvbto$ methods.

\begin{figure}
\centerline{\epsfig{file=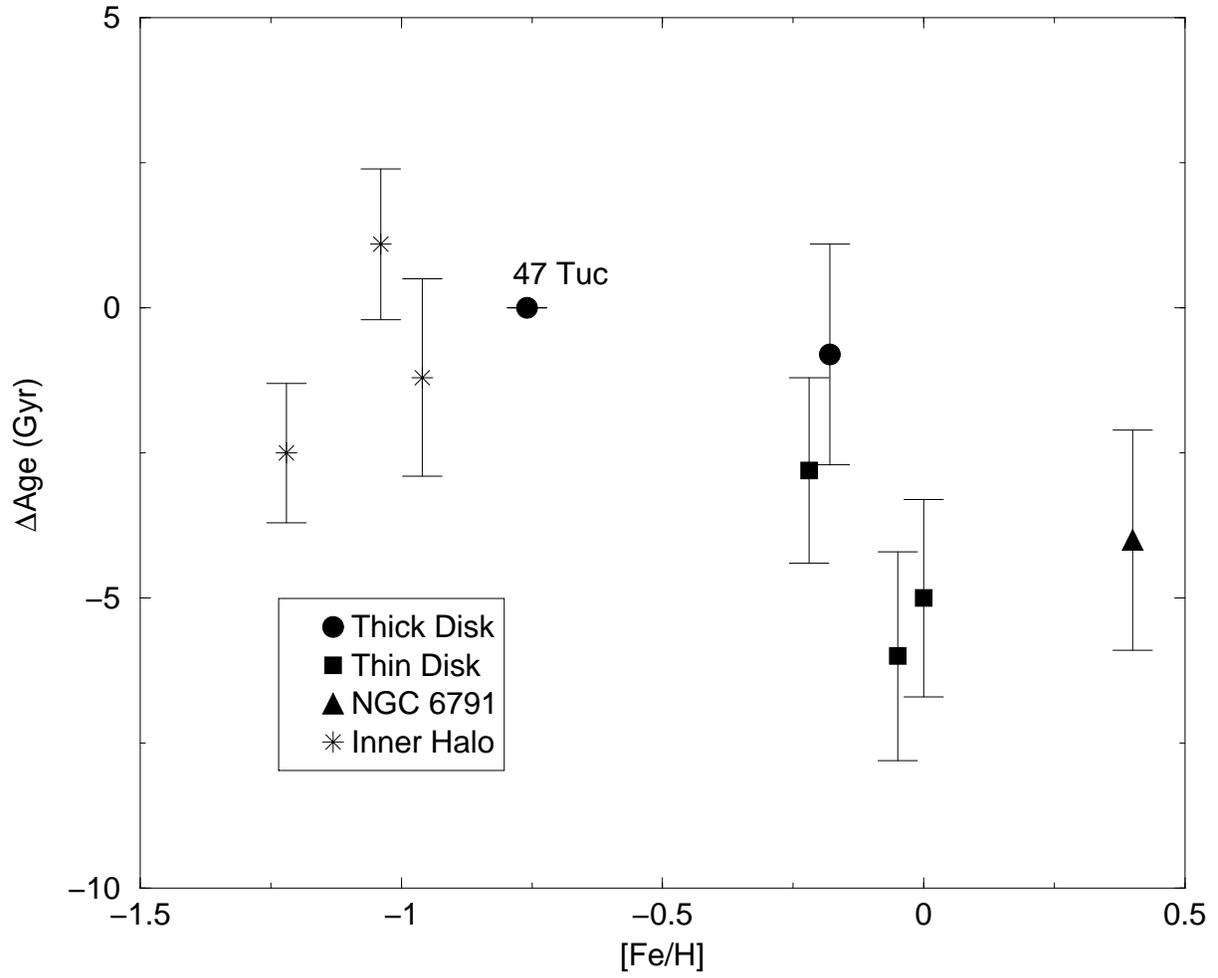,height=16.0cm,angle=270}}
\caption{Ages relative to 47 Tuc as a function of metallicity for the
stars and clusters discussed in this paper.}
\label{figfehage}
\end{figure}

The relative ages determined in this paper indicate that NGC 188 was
formed $6.0\pm 1.8$ Gyr after 47 Tuc.  The oldest solar metallicity
Hipparcos field stars were formed $5.0\pm 1.7$ Gyr after 47 Tuc.  Both
of these results suggest that the solar metallicity stars in the thin
disk were formed signifiantly later than the thick disk.  Simply
averaging these results suggest that solar metallicity stars (in the
solar neighborhood) started to form $5.5\pm 1.2$ Gyr after the thick disk.
In contrast,
the slightly metal-poor ($\feh \approx -0.22$) solar neighborhood
stars were formed $2.8\pm 1.6$ Gyr after 47 Tuc.  As this difference
in age is less than $2\,\sigma$ it is conceivable that thin disk stars
were being born in the solar neighborhood at the same time, or shortly
after thick disk objects like 47 Tuc were forming.
The small relative ages between the thick disk and
inner halo (Table \ref{tabresults}) provide evidence that star formation
in these regions began at roughly the same time.  Figure
\ref{figfehage} shows the relative ages versus metallicity of all the
clusters discussed in this study.

It is interesting to note that the very metal-rich open cluster NGC
6791 is  $4.0\pm1.9$ Gyr younger than 47 Tuc.  This old open cluster
has an orbit which makes its classification as a thin disk object
somewhat problematic.  In particular \citet{Scott95} report that this
cluster has a rotation velocity $\Theta$ which is 60 km/s lower than the thin
disk and it is moving radially outward with $\Pi = 100$ km/s
resulting in a orbit with a large eccentricity. However, this work
assumed a considerably larger distance than was found by CGL (5.3 vs.\
4.2 kpc).  For this reason, we have calculated the kinematic and
orbital parameters of NGC 6791 using the distance found by CGL (4.2
kpc), the radial velocity from \citet{peterson} ($47\pm 1$ km/s) and the
preliminary proper motions from  \citet{cudworth}.  We find the
following:
$\Pi = 88\pm 14 $ km/s; $\Theta = 165\pm 10$ km/s; $W = -6\pm
17$ km/s; $L_z = 1266$ kpc km/s; $R_{apo} = 9.1$ kpc; $R_{per} = 4.2$
kpc; $z_{max} = 0.8$ kpc; $e = 0.37$ and $\Psi = 6.6^{\degr}$.  
The orbit is fairly eccentric with a low rotation velocity and a
reasonably large $z_{max}$ suggesting that NGC 6791 is a thick disk
object. However, 
\citet{Scott95} caution that clusters like NGC 6791 may simply be
in the wings of the kinematic distribution of the thin disk, and its
peculiar orbit is what has allowed it to survive.   

Table \ref{tabresults} includes the median distance of the
stars/clusters from the galacic center ($R_m$) based upon the orbits
of the stars/clusters.  We see that all of the thick disk and thin
disk objects we have discussed in this paper cover a relatively modest
range in their distance from the Galactic center ($5.9 \le R_m \le
10.1$).  Thus, all of these object were born at reasonably similar
Galacto-centric distances at so their ages reflect the star
formation in small region of the galaxy.  The 3 inner halo clusters
discussed by CSA turn out to have a fairly wide range in $R_m$ (2 to
18 kpc).  The fact that the 47 Tuc has a similar age to these clusters
suggests that globular cluster formation was occuring at about the
same time over a wide range of Galacto-centric distances

There have been many proposed scenarios of galactic formation.  These
scenarios include the free-fall collapse of a protogalactic cloud,
originally proposed by \citet{Egg62}, and the merger of independently
formed stellar systems \citep{Sea77,sz78}.  Although no definative
conclusions about galactic formation can be drawn from the results of
this study, some possibilities can be presented.  The similarity in
age between the inner halo and thick disk indicates that if the thick
disk was formed from the gravitational collapse of the halo, this
collapse occurred on short timescales ($<1$ Gyr).  There is a clear age
gap of several Gyr between solar metallicity stars in the nearby thin
disk and the thick disk (47 Tuc).  The  age gap between 
moderately metal-poor thin disk stars and the thick disk is much
smaller, and the present uncertainties in the age determinations
allow for the possibility that the thin disk formed immediately after 
($< 1$ Gyr) after the thick disk and halo.  

\acknowledgments 

We are extremely gratefull to Dana Dinescu for providing us with the
routines to calculate space velocities of the field stars, for
calculating the orbits of the stars and for cross-referencing her
membership list for NGC 188 with the photometry from Sarajedini et al.
We would like to think the anonymous referee and the editor Jim Liebert
whose numerous suggestions led to a considerable improvement in this
paper.  This research was supported by NASA through the NASA/New
Hampshire Space Grant, an LTSA award NAG5-9225 (BC), by a Burke
Research Grant from Dartmouth College (BC) and used the SIMBAD
database, operated by operated at CDS, Strasbourg, France. BC
acknowledges the hospitality of the Aspen Center of Physics, where he
revised the paper.

\clearpage

\begin{deluxetable}{ccccc}
\tablecaption{Isochrone Fit Parameters for NGC 188 \label{tabn188}}
\tablewidth{0pt}
\tablehead{
\colhead{\feh} & 
\colhead{\dmv}&
\colhead{\ebv}&
\colhead{\evi}&
\colhead{Age (Gyr)}
}
\startdata
 $-0.05$ & 11.43 & 0.09 & 0.1125 & 6.5\\
 $-0.15$ & 11.34 & 0.09 & 0.1125 & 7.5\\
 $+0.05$ & 11.48 & 0.09 & 0.1125 & 5.5\\
 $-0.05$ & 11.31 & 0.07 & 0.0875 & 7.5\\
 $-0.05$ & 11.52 & 0.11 & 0.1375 & 5.5\\
 \enddata
\end{deluxetable}

\begin{deluxetable}{rrrrrrrr}
\tablecaption{Solar Neighborhood Stars: Basic Data\label{tabfield}}
\tablewidth{0pt}
\tablehead{
&&&&&
\colhead{$V_{\rm radial}$}&
\colhead{$\mu_{\alpha}\cos \delta$ }&
\colhead{$\mu_{\delta}$ }\\

\colhead{HD} &
\colhead{\feh} & 
\colhead{\bv}&
\colhead{$\pi$ (mas) }&
\colhead{${\rm M_V}$}&
\colhead{(km/s)}&
\colhead{(mas/yr)}&
\colhead{(mas/yr)}\\

}
\startdata
 11007 &$-0.18$ &   0.57 &$36.65\pm  0.7 $ & $3.60\pm  0.04$  &$ -26.5$&$ -166.73$&$  297.35$\\
 15335 &$-0.22$ &   0.59 &$32.48\pm  0.8 $ & $3.45\pm  0.06$  &$  40.3$&$  -65.37$&$   72.52$\\
 32923 &$-0.20$ &   0.66 &$63.02\pm  0.9 $ & $3.91\pm  0.03$  &$  20.3$&$  536.05$&$   18.51$\\
 41330 &$-0.24$ &   0.60 &$37.90\pm  0.8 $ & $4.01\pm  0.05$  &$ -11.8$&$ -124.24$&$ -295.30$\\
 52711 &$-0.15$ &   0.60 &$52.37\pm  0.8 $ & $4.53\pm  0.03$  &$  21.8$&$  155.73$&$ -828.01$\\
 67458 &$-0.24$ &   0.60 &$39.08\pm  0.8 $ & $4.76\pm  0.04$  &$ -17.6$&$  339.59$&$ -354.69$\\
202628 &$-0.14$ &   0.64 &$42.04\pm  0.9 $ & $4.87\pm  0.05$  &$  10.7$&$  242.07$&$   21.98$\\
207129 &$-0.15$ &   0.60 &$63.95\pm  0.8 $ & $4.60\pm  0.03$  &$ -7.0 $&$  165.64$&$ -295.00$\\
210918 &$-0.18$ &   0.65 &$45.19\pm  0.7 $ & $4.51\pm  0.03$  &$-18.0 $&$  570.33$&$ -791.08$\\
 \enddata
\end{deluxetable}

\begin{deluxetable}{rrrrrcrrrrrr}
\tabletypesize{\footnotesize}
\rotate
\tablecaption{Solar Neighborhood Stars: Derived Data\label{tabfield2}}
\tablewidth{0pt}
\tablehead{
&&
\colhead{$\Pi$}&
\colhead{$\Theta$}&
\colhead{$W$} &
\colhead{$L_Z$} &
\colhead{$R_{apo}$} &
\colhead{$R_{per}$} &
\colhead{$z_{max}$} &
&
\colhead{$\Psi$} \\

\colhead{HD} &
\colhead{Age (Gyr)}&
\colhead{(km/s)}&
\colhead{(km/s)}&
\colhead{(km/s)}&
\colhead{(kpc km/s)}&
\colhead{(kpc)}&
\colhead{(kpc)}&
\colhead{(kpc)}&
\colhead{$e$}&
\colhead{(deg)}&
\colhead{Population}\\
}
\startdata 
 11007 &$  6.6\pm 0.5$   &$ -37.2\pm 1.3$&$  250.8\pm 1.2$&$ 48.2\pm  1.1$  &  2006.4 &  11.83& 7.71& 0.84& 0.211&  4.85& Thick Disk?\\
 15335 &$  6.5\pm 0.5$	 &$  14.3\pm 1.5$&$  265.8\pm 1.0$&$ -6.6\pm  1.0$  &  2126.4 &  12.66& 7.96& 0.08& 0.228&  0.44& Thin Disk\\
 32923 &$ 10.0\pm 0.5$	 &$  14.8\pm 0.9$&$  210.1\pm 0.3$&$ 35.9\pm  0.5$  &  1680.8 &   8.20& 7.24& 0.46& 0.063&  3.45& Thin Disk\\
 41330 &$  9.3\pm 0.6$	 &$ -17.4\pm 2.0$&$  208.7\pm 0.6$&$-25.1\pm  0.8$  &  1669.6 &   8.22& 7.05& 0.29& 0.077&  2.15& Thin Disk\\
 52711 &$  6.7\pm 1.0$	 &$   4.5\pm 1.9$&$  156.9\pm 1.2$&$ -2.7\pm  0.6$  &  1255.2 &   8.00& 4.26& 0.02& 0.306&  0.22& Thick Disk\\
 67458 &$  6.5\pm 1.0$	 &$ -72.6\pm 1.1$&$  227.8\pm 0.9$&$ 18.3\pm  0.2$  &  1822.4 &  10.95& 6.64& 0.21& 0.245&  1.41& Thick Disk?\\
202628 &$  6.3\pm 1.9$	 &$   0.5\pm 1.5$&$  235.8\pm 0.1$&$-19.5\pm  1.5$  &  1886.4 &   9.39& 8.00& 0.23& 0.080&  1.51& Thin Disk\\
207129 &$  6.2\pm 1.3$	 &$   2.6\pm 3.2$&$  211.9\pm 0.6$&$  8.1\pm  3.8$  &  1695.2 &   8.01& 7.38& 0.08& 0.041&  0.61& Thin Disk\\
210918 &$ 11.7\pm 1.1$	 &$  37.0\pm 1.3$&$  142.2\pm 1.7$&$ -1.0\pm  1.7$  &  1137.6 &   8.19& 3.60& 0.01& 0.390&  0.09& Thick Disk\\
 \enddata
\end{deluxetable}

\begin{deluxetable}{cccccc}

\tablecaption{Isochrone Fit Parameters for 47 Tuc \label{tab47tuc}}
\tablewidth{0pt}
\tablehead{
\colhead{\feh} & 
\colhead{$[\alpha/{\rm Fe}]$} &
\colhead{\dmv}&
\colhead{\ebv}&
\colhead{\evi}&
\colhead{Age (Gyr)}
}
\startdata
 $-0.80$ & 0.40 & 13.45 & 0.04 & 0.05 & No simult. fit\\
 $-0.70$ & 0.40 & 13.40 & 0.04 & 0.05 & No simult. fit\\
 $-0.95$ & 0.40 & 13.50 & 0.04 & 0.05 & $12.5 \pm 0.5$ \\
 $-0.95$ & 0.40 & 13.28 & 0.03 & 0.0375 & $14 \pm 0.5$ \\
 $-0.95$ & 0.40 & 13.42 & 0.05 & 0.0625 & $11.5 \pm 0.5$ \\
 \enddata
\end{deluxetable}

\begin{deluxetable}{llrcclc}
\tablecaption{Relative Ages between Old Thin, Thick Disk and Halo Stars \label{tabresults}}
\tablewidth{0pt}
\tablehead{
\colhead{Object} &
\colhead{Location}&
\colhead{$\feh$}&
\colhead{$R_m$ (kpc)\tablenotemark{a}} &
\colhead{$\Delta$Age (Gyr)}&
\colhead{Method}&
\colhead{Source}
}
\startdata
47 Tuc      & Thick disk & $-0.71$ & 6.2 &  0 & Iso. fit \& \dvbto & This paper\\
Field Stars & Thick disk & $-0.18$ & 5.9 & $-0.8 \pm 1.9$ & Isochrone  & This paper\\
NGC 6791    & Thick disk? & $+0.40$ & 6.6 & $-4.0 \pm 1.9$ & Isochrone fit & CGL\\
Field Stars & Thin disk & $-0.22$  & 7.6 & $-2.8 \pm 1.6$ & Isochrone  & This paper\\
Field Stars & Thin disk & $0.00$   & 8.0 & $-5.0 \pm 1.7$ & Isochrone  & This paper\\
NGC 188     & Thin disk & $-0.05$  & 10.1\phn & $-6.0 \pm 1.8$ & Isochrone fit & This paper\\
M107        & Inner halo & $-1.04$ & 2.8 & $+1.1 \pm1.3$ & \dvbto & CSA\\
NGC 1851    & Inner halo & $-1.22$ & 18.0\phn & $-2.5 \pm1.2$ & \dvbto & CSA\\
NGC 6652    & Inner halo & $-0.96$ & (2.0) & $-1.2 \pm1.7$ & \dvbto & CSA\\
 \enddata
\tablenotetext{a}{ 
$R_m = (R_{apo} + R_{per})/2$ (median distance from the Galactic
center), determined from the stellar orbits (\S \ref{fieldstars} for
the field stars; \S \ref{secsumm} for NGC 6791; 
\citet{dana} for the globular clusters; and \citet{carraro} for NGC
188).  
The numbers in parathesis for NGC 6652 is its $R_{GC}$ value, as 
orbital information does not exist for this cluster.}

\end{deluxetable}

\clearpage

\end{document}